# The Asset Price Channel of Monetary Policy: Evidence from Regional Stock-Market Developments in the Successor States of Former Yugoslavia

**Stefan Tanevski**
stefan.tanevski@uacs.edu.mk
**University American College Skopje**

**Abstract**

The aim of this study is to empirically investigate the existence of a sectoral asset price channel of monetary policy in the region of the six republics of former Yugoslavia. The study constructs sectoral indices for the entire region, building on the idea that one regional stock exchange may provide more efficiency for the listed companies in the region, while monetary policy relevance for it may be sector-specific. We employ panel vector autoregressive model to observe impulse responses of sectoral indices to innovations in monetary policy, while then disentangle the long- from the short-run relationships per index through a Pooled Mean Group estimation. Overall, we document presence of the asset price channel in the finance and telecom sectors, likely driven by the established multinational corporate networks fostering sub-market regionalization. Yet, this is not the case for the manufacturing and electricity sectors, which may imply that local stock markets are yet too fragmented and space for a more efficient regional stock market, either in the true sense of the word or, more realistically, though enhanced regional cooperation of the stock exchanges certainly exists.

## 1. Introduction

Following the latest structural shocks i.e., covid-pandemic, supply chain disruptions and energy crisis, central bank authorities have been forced to react after decade long zero-policy rate in an attempt to restore price-stability. The repercussions of these policy adjustments (delayed signaling followed by aggressive tightening), have captured academic attention (Thorbecke, 2023). Yet, despite the initial perception of the structural shocks having a transitory impact on prices, they have led to a reassessment of the extent to which monetary policy affects macroeconomic variables, including asset prices in the post-pandemic period.

Structural shifts in the past two decades, such as financial openness characterized by relaxing barriers to cross-border capital flows and financial liberalization, have played a significant role in emerging economies where bank-based financing has traditionally been dominant (IMF 2021). These shifts, coupled with other market-oriented reforms, have paved the way for transmission channels previously dormant. The introduction of an inflation targeting regime at the beginning of the century, with interest rates as the main operating target, has notably improved monetary policy pass-through in many emerging economies (Mehrotra and Schanz, 2020). This transformation has not only allowed for increased liquidity but has also fostered development in domestic financial markets, particularly in securities markets (Andreasen and Valenzuela, 2016; Tongurai and Vithessonthi, 2022; Lee and Chou, 2022).

Recent work has re-examined the issue of monetary policy, inflation and asset prices following the pandemic. Gagliardone and Gertler (2023) demonstrate that easy monetary policy (i.e., delayed response) coupled with oil shocks propagated the recent surge in inflation even when accommodating for demand shocks and labor market tightness. In a similar vein, Bernanke and Blanchard (2023) attribute price increases to supply-chain constraints and slow adjustment of monetary policy. Besides adjusting the policy rate, central banks employed alternative mechanisms i.e., Quantitative Easing (QE). Such interventions are found to impact risk premia on assets (Delgado and Gravelle, 2023). In this context, a recent work by Kashyap and Stein (2022) revisits how central bank authorities shape market sentiment. They find extensive evidence of reduced risk premia during monetary policy easing. This finding extends to euro and non-euro area countries (IMF, 2023). Although at a different magnitude, the sample of euro-area countries i.e., Greece, Germany and Italy exhibit higher transmission of shocks via stock exchanges relative to non-euro area countries.

Moreover, local capital markets in emerging economies have undergone multiple-stages of development (Woolridge, 2020). In some large emerging economies such as China, this process has led to a rapid increase in the effectiveness of the asset price channel, as a result of increased proportion of securities and real estate (Ajaz et al., 2017; Li et al., 2021). Existence of an effective asset price channel has also been documented in other regions of Asia (Kurniawan and Astuti, 2023) and in fast-transitioning countries in Central and Eastern Europe (CEE) and Eastern Europe (EE) (Stoica et al., 2013), while evidence for some African countries is ambiguous (Tchereni et al., 2022; Belmouss et al., 2023). Nonetheless, inferring general conclusions on the transmission effectiveness of monetary policy in emerging countries is difficult as outcomes are quite conditional on country and region specifics.

Yet, in the republics of the former Yugoslavia, where market-oriented reforms and structural shifts began roughly three decades ago, such reforms related to openness, euroization, EU-membership, can play instrumental role in the attractiveness of financial markets. This, in turn can have positive implications for the region's development and for its transmission of monetary policy. An important aspect in that regard has been the intention to regionalize the local stock markets, ideally through one operational Balkan Stock Exchange, while more realistically as enhanced cooperation of and cross-listing of local companies on the other stock exchanges in the countries of former Yugoslavia. The economic integration of the region has been present in the last decade or two at the least: for example, all of the republics of former Yugoslavia (and some other) form a trade block known as CEFTA, the EU-supported Berlin process fosters unification of many economic processes, while the recent Open Balkan Initiative, despite smaller in scope, adds to the process. In parallel, the financial integration is at least partially ensured through the presence of European large bank in the bank-dominated financial markets of the region, supported by similar emerging trends in insurance, pension funds and investment funds. Multinational corporations are likewise present and strengthen the mutual ties of the region and those of the region with the EU and the rest of the world, common multinationals being present in some sectors like the telecommunications or hotels.

To explore this issue, the paper examines the **operation of the asset price channel of monetary policy** within the **context of regional stock-market developments** among the successor states of former Yugoslavia. Rather than analyzing the relationship between monetary policy and stock prices within individual national markets, the study adopts a **sector-based regional perspective**, allowing the effects of monetary shocks to be viewed through both **domestic and cross-border transmission mechanisms**. To our knowledge, this perspective has not previously been applied to these economies.

The analysis relies on **self-constructed sectoral stock-market indices** aggregating data from listed companies in North Macedonia, Serbia, Montenegro, Bosnia and Herzegovina, Croatia, and Slovenia. This design reflects the view that **stronger financial linkages whether through formal exchange cooperation or gradual harmonization, may improve market liquidity and enhance the consistency of monetary transmission**. At the same time, the framework recognizes that **countries differ in their monetary regimes: Slovenia and Bosnia and Herzegovina**, being tied to the **Eurosystem**, are directly influenced by **Euribor-based monetary conditions**, whereas **non-Eurozone economies** maintain domestic policy rates yet remain **exposed to external liquidity and exchange-rate constraints**.

In what follows, the paper outlines the main channels through which monetary policy affects real and financial variables, with particular attention to the **asset price channel** and its evolution in the literature. Section 3 describes the data and methodology, Section 4 presents the empirical results and discussion, and Section 5 concludes.

### 1.1. Stylized Section: Stock Markets in Western Balkans (2005-2023)

The structure and performance of stock markets across the Western Balkans remain among the key determinants of how monetary impulses propagate to asset prices. Despite notable institutional progress since the mid-2000s, equity markets in Croatia, Slovenia, Serbia, Bosnia and Herzegovina, and North Macedonia continue to exhibit features typical of small, bank-dominated, and thinly traded markets.

#### Market Size and Capitalization

Across the region, market capitalization expanded rapidly in the pre-2008 period, reflecting post-privatization listings and buoyant regional liquidity, but stagnated thereafter. As shown in *Figure 1*, capitalization rose from roughly €1–5 billion in the smaller markets (Bosnia, North Macedonia, Serbia) and over €10 billion in Croatia by 2007, before moderating after the global financial crisis. By 2020, capitalization stood around €18 billion in Croatia, €6 billion in Slovenia, and under €3 billion in most others, with ratios to GDP rarely exceeding 30 %

**Figure 1. Market Capitalization (million EUR, 2005–2023)**

**Source: Author's Calculations**

In relative terms, capitalization-to-GDP shown in **Table 1**, peaked at 50 % in Croatia and 34 % in Serbia around 2010, but has since declined to below 20 % in most markets. The exceptions are North Macedonia, where capitalization reached 31 % of GDP in 2020, supported by a few large cross-listed firms, and Slovenia, which maintains steady though modest capitalization around 14 % of GDP.

Table 1. Market Capitalization to GDP

| Market Capitalization to GDP | | | | |
|---|---|---|---|---|
| **Country** | **2005** | **2010** | **2015** | **2020** |
| **Croatia** | 27% | 50% | 37% | 39% |
| **Slovenia** | 17% | 20% | 13% | 14% |
| **Macedonia** | 14% | 30% | 18% | 31% |
| **Serbia** | 16% | 34% | 15% | 7% |
| **Bosnia** | 14% | 17% | 15% | 13% |
| **Source: Author's Calcilations** | | | | |

### Liquidity and Turnover

Liquidity indicators confirm persistent market shallowness. Aggregate turnover volumes (Figure 2) show that after peaking during 2006–2007, trading activity collapsed and never recovered to pre-crisis levels. Croatia, the region's largest exchange, saw turnover fall from €9 billion in 2007 to €0.4 billion in 2023. Similar contractions occurred in Slovenia and Serbia.

**Figure 2.** Turnover (million EUR, 2005–2023)

Source: Author's Calculations

*Note: The figure reports turnover values for selected benchmark years (2005, 2006, 2007, 2008, 2009, 2016, 2017, 2018, and 2020)*

The turnover-to-GDP ratio further illustrates this decline: between 2005 and 2020, it dropped from 37 % to 2 % in Croatia, 22 % to 6 % in Slovenia, and 10 % to 1–3 % in the remaining markets. Even at their peak, turnover ratios across the Western Balkans rarely exceeded 10 %, compared to 30–50 % in Central-European or Baltic peers.

### Breadth and Depth

Measured by the number of listed firms per 10 000 inhabitants, market breadth has narrowed over time. Table 1 shows that the average fell from roughly 0.5 in 2005 to below 0.3 by 2020, with the steepest contractions in Croatia and Slovenia. Bosnia and Herzegovina remains an outlier, with a relatively high number of listings due to its post-war privatization wave, but turnover remains negligible, implying very low secondary-market activity.

Table 1. Summary Indicators of Western Balkan Stock Markets, 2005–2023

| Indicator (Average, 2005–2023) | Croatia | Slovenia | North Macedonia | Serbia | Bosnia |
|---|---|---|---|---|---|
| **Market Cap (% GDP, avg)** | 35% | 15% | 23% | 18% | 14% |
| **Turnover (% GDP, avg)** | 10% | 7% | 4% | 3% | 3% |
| **Firms per 10 000 inhabitants** | 0.4 | 0.2 | 0.4 | 0.1 | 2.0 |
| **Source: Author's Calcilations** | | | | | |

In this context, the stylized evidence underscores that market depth, not mere existence, determines the strength of monetary-policy transmission through equity prices. The persistence of low capitalization, thin trading, and narrow sectoral composition suggests that the asset-price channel in the region operates only partially consistent with the empirical results presented later in this paper.

## 2. Monetary policy transmission, with focus on the asset price channel

Monetary policy influences real economic activity through several well-established transmission mechanisms. The traditional view emphasizes the **interest-rate channel**, where policy-induced changes in short-term rates affect the cost of capital and investment spending (Bernanke and Gertler, 1995; Taylor, 1995). This mechanism operates primarily through borrowing costs and remains a central feature of conventional models. Over time, scholars have proposed extensions to account for frictions in credit markets, including the **credit and balance-sheet channels**, which amplify monetary effects through variations in firms' collateral values, external-finance premia, and credit availability (Bernanke and Gertler, 1989; Mishkin, 1996). Although these channels remain important, their relative strength has diminished in increasingly market-based and financially liberalized economies.

In contrast, the **asset price channel** provides a distinct perspective on monetary transmission, focusing on how monetary impulses affect equity valuations and, through them, investment and consumption decisions. Given the growing though still shallow capital markets of the Western Balkan economies, this mechanism is particularly relevant for understanding how monetary signals propagate in environments characterized by partial financial integration and evolving stock exchanges.

***The Asset Price Channel*** as a transmission mechanism operates through the stock market. Literature proposes two mechanisms operating under the asset price channel i.e., Tobin's Q and Wealth Effect.

Initially presented in Tobin's (1961), the first mechanism implies that monetary policy shocks to the aggregate economy transmit through equities via investment spending. In Tobin's (1961), Tobin's q is defined as the difference between market value of firms and their replacement cost of capital. In this context, firms characterized by high Tobin's q may obtain cheap capital (relative to their market value) with a small issuance of equity, leading to a rise in investment spending.

The convergence between monetary policy and equities can be explained from two different points of view. First, a monetarist's view suggests that a rise in money supply induces a change in spending thereby increasing demand for equities. Second, a Keynesian's view, which operates through interest rates, suggests that during periods of relaxed monetary policy bonds become less attractive than equities, leading to a rise in stock prices. Mishkin (1996) advocates that when both views are combined, there is a rise in stock prices, leading to higher Tobin's q and ultimately higher investment spending. From there, we derive the following schematic:

*M ↑ ⇒ Pe ↑⇒ Tobin's q ↑ ⇒ I ↑ ⇒ Y↑*

Crowder (2006), for the second mechanism, presents a mathematical model that captures two ways in which monetary policy affects stock returns and therefore the aggregate economy:

$$P_{t+1} = E_t \left[\Sigma\left(\frac{1}{1+R_t}\right)^j D_{t+j}\right] + E_t\left[\left(\frac{1}{1+R_t}\right)^K P_{t+k}\right]$$

Whereby: ($P_{t+1}$) is the present value of a share, $E_t$ represents the expectations operator based on available information at given time t, Rt is the rate of return applied by market participants, ($D_{t+j}$) is the value of discounted future cash flows, while K represents the holding period or time horizon. A change in the central bank's base rate influences expected returns and participants' expectations, leading them to revalue equites. The effects are depicted by the following schematic:

*M ↑ ⇒ aggregate output ↑ ⇒ firm's profitability ↑ ⇒ Pe ↑*

First, as a result of positive policy shocks to the aggregate economy, economic activity will increase, in turn raising firm's earnings and ultimately stock prices. Second, monetary policy shocks can induce changes in stock prices directly by altering discount rates and thereby expected future cash flows (Crowder, 2006).

Sellin (2001) suggests that a period of tightening should induce a decrease in investors' financial wealth, as a result of lower expected future cash flows or higher discount rate, in turn reducing private investment and consumption expenditure. In Sellin's (2001) and Mishkin's (1996) jargon, this is the 'wealth effect'. The model is depicted by the following schematic:

$$M\downarrow \Rightarrow i\uparrow \Rightarrow Pe\downarrow \Rightarrow \textit{financial wealth}\downarrow \Rightarrow \textit{consumption \& investment}\downarrow \Rightarrow \mathrm{Y}\downarrow$$

The 'wealth effect' assumes that a monetary tightening induces a fall in stock prices by impacting expected future cash flows or the discount rate used by market participants. Fuhrer (1996) reveals that many long-term rates follow closely the changes in the short-term base bank rate. Substantial amount of literature supports the notion that equity returns fall during negative monetary policy shocks (Kuttner, 2001; Rigobon and Sack, 2003; Bernanke and Kuttner, 2005; Ehrman and Fratzscher, 2004). However, monetary policy may be endogenous in that stock market movements can influence the aggregate economy by exerting pronounced volatility and boom-bust cycles and thereby be of consideration in the policy design and, to an extent, a ground for central bank reaction (Crowder, 2006).

The issue of whether central bank authorities should adopt a reactive or proactive stance towards asset price misalignments has been a long-standing debate in economic circles. The reactive approach posits that an ex-post reaction, or a "wait-and-see" approach, is preferred, where a reaction is deemed necessary only if price misalignments reverse and have an impact on price stability and output. To this end, Bernanke and Gertler (2000) suggest that the implementation of a flexible inflation targeting regime would be an effective tool to stabilize output and eliminate misalignments. On the other hand, the proactive approach advocates for authorities to respond to developing price misalignments, i.e., market bubbles. Cecchetti et al. (2000) propose that significant deviations in asset prices may cause price and labor instability that can be further exacerbated when the misalignments are diminished. Therefore, if the asset price deviations originate from the asset markets themselves, a reaction is imminent. Such policies should aim to attenuate disturbances on the real economy, and a preemptive policy may ultimately limit the accumulation of such misalignments and prevent economic imbalances, which would reduce the need for medium-term corrections to price stability and output.

**Review of the empirical literature on the asset price channel**

The relationship of monetary policy and stock prices evaluated through the asset price channel has been extensive. Hence, we attempt to make an overview of the existing literature on this topic while capturing all periods starting from 1960s.

*Early studies,* focused on examining the impact of an increased money supply on stock prices. A body of evidence (Sprinkel, 1964; Keran, 1971; Hamburger and Kochin, 1972) arrived at conclusions that contradicted Fama's (1970) Efficient

Market Theory. These studies suggested that changes in stock prices result from monetary changes, challenging the notion that stock prices purely reflect all available information. However, Pesando (1974) raised concerns about the established relationship in Sprinkel's (1964) and Keran's (1971) studies due to the presence of serial correlation in individual series. Additionally, most earlier studies overlooked the concept of 'information lag.'

Rozeff (1974) took a step further by testing the Efficient Market Hypothesis against "The Monetary Portfolio Model", which views money as an asset. Developed by Friedman and Schwartz (1963), the Monetary Portfolio Model implies that any shocks to the money supply will result in portfolio rebalancing as investors wish to reestablish desired money holdings. Rozeff (1974) critiques past studies for failing to distinguish that monetary shocks affect stocks with a lag i.e., they employ ex-post data of money supply, assuming zero lag in publication.

Building on Rozeff's work, Rogalski and Vinso (1977) improved the model by synchronizing data intervals, capturing money supply data at the same intervals as stock return data, and accounting for autocorrelation. Their enhanced model supported Rozeff's findings and confirmed their robustness. The research demonstrated that causality and directionality exist from stock prices to money supply and vice versa, suggesting the potential presence of endogeneity bias in earlier studies.

The presence of endogeneity bias in earlier studies prompted a shift toward a different approach - event studies. Cornel (1983) advocates for the announcement approach, citing two key reasons. Firstly, with respect to the Federal Reserve's announcement, the reported money stock figure is independent of Fed policy or asset prices, as it pertains to the week ending nine days earlier. Any correlation observed between money supply announcements and asset prices suggests a directionality from announcements to asset prices. Secondly, working with announcement dates allows for the use of higher frequency periods and a more powerful observation interval, addressing a limitation of previous studies. Cornel (1983) acknowledges that longer-term intervals might be preferable unless the impact of monetary shocks on asset prices persists for a quarter or longer.

Among the first to take stock of the issue of announcements are Berkman (1978) and Lynge (1981). They find evidence of a negative relationship between money supply announcement and stock prices. Berkman (1978) however, is the first to differentiate between anticipated and unanticipated changes in money supply, suggesting that reaction is imminent only when changes are unanticipated. Lynge's work, on the other hand, does not bear directly on the Efficient Market Hypothesis. Subsequently, Pearce and Roley (1983) reexamined previous evidence using weekly data obtained from survey expectation measure for the period 1977 – 1982, estimating the following model whereby,

$$\Delta P_t = a + b(\Delta M_t^a - \Delta M_t^e) + \varepsilon_t,$$

Where the percentage change of stock price is expressed as a function of the difference between change in announced money stock and expected money stock. Obtaining a negative estimate for 'b', Pearce and Roley's (1983) work corroborated previous findings. A consensus on the impact and effect of weekly money stock announcement (i.e. positive surprises) has also been investigated and corroborated in Urich and Wachtel (1981) and Roley (1982)].

A number of competing hypotheses have been presented in order to explain the asset price reaction to announcements. *The expected inflation hypothesis* is that announcement of unanticipated increase in money stock, raises participants' expectations of inflation, leading to higher rates (Fisher effect). Based on the discounted cash flow model, agents will revise their expected earnings, in turn causing stock prices to fall. *The Keynesian hypothesis* as based on the sticky price model suggests that money supply announcements will have effect only if they alter expectations about future monetary policy. In this context, positive shocks will precede agents to expect a tightening of monetary policy, lowering stock prices. *The real activity hypothesis* proclaims that announcement of bigger money supply provides information regarding the future money demand. For this link to operate money demand is assumed to depend on expected future output, which infers higher expected future cash flows and thus stock prices. Cornel (1983) addresses the four hypotheses. He finds that when studying them simultaneously at least three can be rejected. In this context, Hafer (1986) also posits that in most cases these hypotheses are not fulfilled when evaluated.

Extending their previous work, Pearce and Roley (1985) investigated effects of economic news i.e., changes in money, inflation and output for the period 1979-1982. They concluded that after a change in Fed's methodology, money announcement surprises have significant negative effect over stock prices. Hafer (1986) supports this notion, indicating that unanticipated changes in money negatively affect stock prices, with no significant difference between the three periods. In contrast, Hardouvelis (1987) finds significance only for the two subperiods (79-82/82-84), demonstrating that monetary variables (money supply) bear significance while non-monetary variables (output) do not—a result that aligns with Pearce and Roley's (1985). They establish that the response of stock prices to macroeconomic news is conditional upon the economic environment. Favorable news about future output in an overheated economy is linked to decreased stock prices, driven mainly by adjustments in expected cash flows rather than alterations in discount rates. Conversely, unfavorable news is associated with an increase in stock prices. These findings provide a rationale for the lack of significance observed in prior research.

The initial evidence supporting the idea that discount rate announcements impact stock prices came from Waud (1970). In similar vein, Roley and Troll, (1984), Smirlock and Yawitz, (1985,) Pearce and Roley , (1985) confirm this, ndicating a significant adverse influence arising mainly from the unanticipated portion of discount rates post-1979 period. Yet, following Santomeros' (1983) suggestion that discount rates respond to market rates with a lag reveals two drawbacks. First, if stock prices react to changes in discount rates, it implies market inefficiency, as even changes aligned with policy objectives are already incorporated into the market. This challenges the explanation of the announcement effect assuming market efficiency. Second, using changes in discount rates as regressors introduces endogeneity bias and therefore misinterpretation of endogenous changes as monetary policy shifts. Subsequently, Smirlock and Yawitz (1985) address this by decomposing the discount rate into 'technical' (endogenous or expected) and 'non-technical' components, the latter conveying information on monetary policy and find evidence of negative effects of announcements only for the period post 1979, yet explicitly for the non-technical components, reconciling the findings of Pearce and Roley (1985). Other prominent articles are those of Hafer (1986) and Hardouvelis (1987).

However, subsequent to Sims' (1980) this issue has been addressed in terms of vector autoregressive models. Prominent articles include, Thorbecke (1997), Patellis (1997), Lastrapes (2001), Cassola Morana (2004). Thorbecke (1997) examines the relationship on stock prices and monetary policy shocks (measured as orthogonalized innovations in the federal funds rate) concluding i) positive and significant effect on stock returns; ii) presence of firm size effect (in line with Gertler and Gilchrist, 1994). Patellis (1997) finds that monetary policy indicators are significant predictors of excess stock returns, which is justified by the propagation mechanism (Bernanke and Gertler, 1988) and credit view (Bernanke and Gertler, 1995) explained in previous sections.

Hitherto, the presented literature is mainly about advanced economies; (Patellis, 1997; Thorbecke, 1997; Lastrapes, 2001; Cassola Morana, 2004; Bernanke and Kuttner, 2005 and Gal´ı and Gambetti, 2015; Thorbecke, 2023; IMF, 2023) and has addressed the issue from an aggregated perspective. However, pertinent literature from developing economies is emerging: Jarocinski (2010), Yoshino et al. (2014), Stoica et al. (2014), Grabowski and Grabowska (2021), although it remains at an aggregate level.

## 3. Data and Methodology

In achieving the objective of our paper and aligning with the literature on transmission channels related to monetary models, we employ the index of industrial production (*ip)* as a proxy for the real economy; the central bank foreign exchange reserves changes (*err)*, to account for external constraints faced by small open and partially euroized economies, we incorporate **changes in central-bank foreign-exchange reserves**. This variable reflects the balance-of-payments position and the degree of exchange-rate stabilization, both of which influence domestic liquidity and the transmission of monetary impulses to asset prices.

Recent evidence confirms that **Euribor itself embodies a credit-risk premium**, as the **Euribor–Overnight Index Swap (OIS) spread** largely reflects the credit quality of the contributing panel banks and transmits to the **financing costs of the real sector**, thereby moderating the pass-through of monetary policy shocks (Ercegovac, Šestanović & Pečarić 2025). This reinforces the interpretation of Euribor as the operative rate faced by borrowers, while acknowledging that variations in the Euribor–OIS spread may create short-term frictions in the policy-rate channel.

Incorporating this consideration allows the specification to capture both the **domestic** and **external** dimensions of monetary-policy transmission that characterize the region's hybrid regimes.

To this set of variables, we would like to add a variable about stock prices. The usual way in the literature is to take aggregate stock market indices. We instead propose alternative approach which disaggregates by sector but aggregates at the regional level. The idea is to understand if domestic monetary policy, which directly or indirectly depends on the monetary policy of the European Central Bank could entice some movements in the sectoral sentiment at the regional level. We first identify the listed companies of most liquid regional stock indices[1] and categorize them by sectors in accordance with the NACE Rev.2 European Classification i.e., Manufacturing, Information and telecommunications, Finance and insurance activities, and Electricity, gas, steam and air conditioning supply. The four indices are denoted with $index_{kjt,}$ whereby *k* takes a value from 1 to 4.

Using data on company market capitalization ($\text{market cap}_{jt}$) and stock prices ($\text{stock price}_{jt}$) over time ($_t$), we estimate the weight of each $\text{company}_{jt}$ in the total sectoral market capitalization. Then we calculate market capitalisation weighted prices to obtain initial index values ($Index_0$) of 100, in order to normalize the index. In periods following (t>0), the index value ($Index_{jt}$) is estimated as $Index_{t-1}$ x $\frac{\Sigma[Stock\ Price jt]}{\Sigma[Stock\ Price jt-1]}$ x Weighted Market $Capitalization_{jt}$. In such manner, we ensure consistent starting value while subsequent index values (t>0) capture the sector related dynamics in a richer dynamics than considering the country level.

[1] Some companies were omitted due to data constraints. The full list of companies included in index construction is provided in Appendix.

The construct secures field for application of a Panel Vector Autoregressive Model or PVAR, which finds support in literature when employed in analyzing transmission of shocks across time (Canova and Ciccarelli, 2013). The PVAR model could be articulated as follows:

$$Y_{it}=A_0+A_1 Y_{it-1}+\ldots+A_p Y_{it-p}+\alpha_i+\varepsilon_{it}, \quad (1)$$

where, $Y_{it}$ is a (8x8) vector of macroeconomic variables ordered as: $X_t = [irs_{jt}, err_{jt}, index_{kjt}, ip_{jt}, \pi_i]$; $A_0$, $A_1$,..., are matrices of coefficients associated with lagged values, representing autoregressive (AR) terms; $Y_{it-1}$, $Yi_{t-2}$,...,$Y_{it}$ are lagged values of the vector of time series variables up to order $p$; $\alpha_i$ is a vector of individual-specific effects or fixed effects and $\varepsilon_{it}$ is a vector of error terms.

The matrix specification allows for contemporaneous relationship between the variables represented by the diagonal (1) which are set to zero, indicating that the value of each variable is influenced by its own past values and the past values of variables that appear earlier in the ordering.

| | $irst_{jt}$ | $err_{jt}$ | $indextelecom_{jt}$ | $indexman_{jt}$ | $indexelec_{jt}$ | $indexfin_{jt}$ | $ip_{jt}$ | $\pi_{jt}$ |
|---|---|---|---|---|---|---|---|---|
| $irst_{jt}$ | 1 | 0 | 0 | 0 | 0 | 0 | 0 | 0 |
| $err_{jt}$ | 0 | 1 | 0 | 0 | 0 | 0 | 0 | 0 |
| $indextelecom_{jt}$ | 0 | 0 | 1 | 0 | 0 | 0 | 0 | 0 |
| $indexman_{jt}$ | 0 | 0 | 0 | 1 | 0 | 0 | 0 | 0 |
| $indexelec_{jt}$ | 0 | 0 | 0 | 0 | 1 | 0 | 0 | 0 |
| $indexfin_{jt}$ | 0 | 0 | 0 | 0 | 0 | 1 | 0 | 0 |
| $ip_{jt}$ | 0 | 0 | 0 | 0 | 0 | 0 | 1 | 0 |
| $\pi_{jt}$ | 0 | 0 | 0 | 0 | 0 | 0 | 0 | 1 |

A common aspect of our and the papers employing VAR methodology in assessing the effects of monetary policy on stock market is the identification, i.e. Cholesky decomposition, which works so that when the $P$ matrix is lower triangular, we obtain a recursive model, implying that a variable i.e., stock returns, is conditional contemporaneously on the innovations to preceding variables, while subsequent variables react to it with a lag. This implies that structural shocks are recovered using short-run restrictions i.e., due to identification choice, they fail to address two important aspects: interdependent simultaneity and reverse causality. In this context, considering the dynamics of financial markets where prices are fast to react and economic agents influence each other (i.e. central bank and the general public), disallowing "simultaneity" would affect such a scenario, as demonstrated by Bjørnland and Leitimo (2009).

Considering that the PVAR model hinges on the prerequisite of variable stationarity, it inevitably involves a trade-off by foregoing valuable long-run information on the co-movement among variables. To address this limitation we continue by employing a panel vector error-correction model (VECM) framework i.e., Pooled Mean Group model with data where the underlying regressors follow integrated process of order I(1), as follows:

$$\Delta Y_{it} = \alpha + \beta(Y_{it-1} - \bar{Y}_i) + \Sigma_{j=1}^{p} \gamma_j \Delta Y_{it-j} + \delta X_{it} + \epsilon_{it,} \quad (2)$$

Where $\Delta$ represents the first difference operator, $Y_{it}$ is the dependent variable for unit i, at time t, $\bar{Y}_i$is the mean of $Y_i$, $X_{it}$ represents exogenous variables, and $\epsilon_{it}$ is the error term. The term $\beta(Y_{it-1} - \bar{Y}_i)$ captures the short-run dynamics, while the lagged differences $\Delta Y_{it-j}$ capture the long-run dynamics. Employing PMG assumes homogenous coefficients across cross-sectional units; while estimating a common set of coefficients for both short- and long-term parameters. Under the assumption of stock market regionalization, this would imply that i) effects of regionalization are expected to be similar across regions, and ii) effects of increased market liquidity, access to capital and overall efficiency would have consistent impact.

Monthly time series data ranging from January 2010 to September 2023 have been used. Domestic three-month money market interest rates and central bank reserves come from central banks' website. Industrial production is obtained from the official State Statistical Offices of each country, and the EURIBOR three-month interest rate is provided by the European Money Markets Institute. Data on stock prices for the purpose of index construction have been sourced from regional stock exchanges, while data on shares outstanding to estimate market capitalization has been obtained from the annual reports of the companies.

Additionally, as a robustness check, reported in **Appendix 2** the analysis is re-estimated using the Mean Group (MG) estimator, which relaxes the long-run homogeneity constraint imposed by the PMG model.

## 4. Results and discussion

### 4.1 Unit roots and model stability

In establishing the stability of our PVAR model, we follow a systematic approach. Frist, we conduct the Fisher-Augmented Dickey-Fuller test on the panel dataset presented in Table (1) to check for unit roots. Variables in levels do not satisfy the stationarity condition. To achieve stationarity, we employ first-order differencing for all variables. Then, we proceed to perform lag reduction and diagnostic tests which ensure that at two (2) lags our model satisfies basic autoregressive assumptions i.e., no eigenvalues greater than 1 – inverse roots are outside of the unit circle. Refer to Table (2) and Figure (1) for the lag choice and stability condition, respectively.[2]

Table 1. Stationarity. Fisher-Augmented Dickey-Fuller

| *Variable name* | *Variable abbreviation* | *Variables in their levels* | *Variables in first differences* |
|---|---|---|---|
| | *Ho: The variable contains a panel unit root (p-values)* | | |
| **Policy interest rate** | $irs_{jt}$ | 0,2163 | 0,0000 |
| **Central bank reserves** | $err_{jt}$ | 0,9964 | 0,0000 |
| **Index of Industrial Production** | $ip_{jt}$ | 0,2916 | 0,0000 |
| **Consumer Price Index** | $\pi_{jt}$ | 1,0000 | 0,0000 |
| **Stock-market index of Telecom** | $indextelecom_{jt}$ | 0,5990 | 0,0000 |
| **Stock-market index of Manufacturing** | $indexman_{jt}$ | 1,0000 | 0,0000 |
| **Stock-market index of Electricity** | $indexelec_{jt}$ | 0,0477 | 0,0000 |
| **Stock-market index of Finance and Insurance Activities** | $indexfin_{jt}$ | 0,9999 | 0,0000 |
| ***Source: Author's calculations.*** | | | |

Table 2. Lag length selection

| **lag** | MBIC | MAIC | MQIC |
|---|---|---|---|
| **1** | -850.386* | 36.6672 | -305.133* |
| **2** | -637.884 | -46.5147* | -274.381 |
| **3** | -327.318 | -31.6339 | -145.567 |
| **4** | . | . | . |
| ***Source: Author's calculations.*** | | | |

[2] While we base our lag length selection on the Akaike Information Criterion, it is worth noting the PVAR model and the results remain robust even if one lag is selected as advised by MBIC or MQIC criteria.

Figure 1. Eigenvalue stability condition

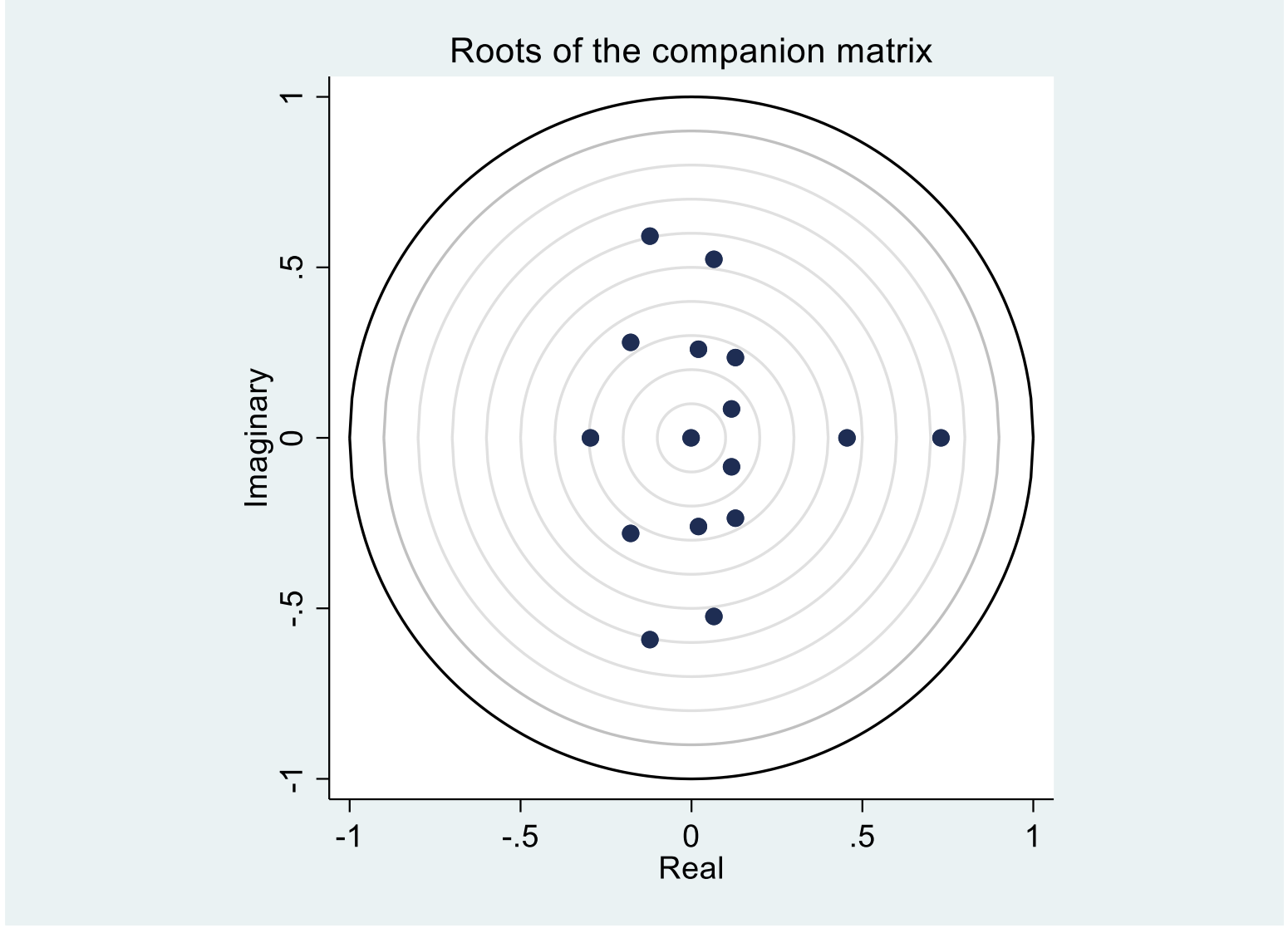

*Source: Author's calculations.*

## 4.2 Impulse response functions

In estimating the short-run dynamics between monetary policy and sectoral stock prices, we ground our analysis on panel VAR Impulse Response Functions presented in Figure (3). Figure (3) presents a reaction of sectoral stock indices to a one percentage point of a monetary policy shock.

Figure 3. IRFs. Monetary policy shocks and sectoral response.

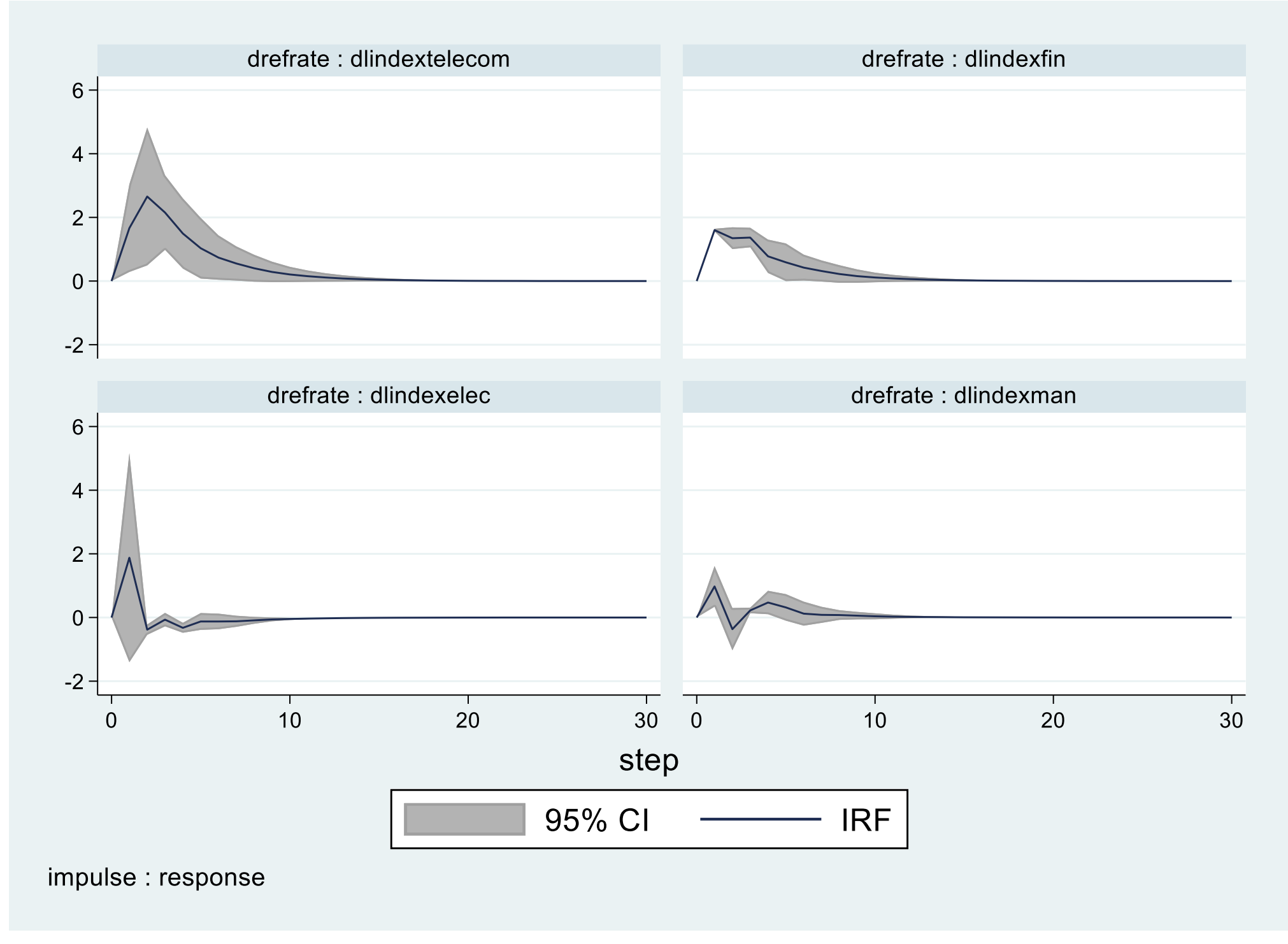

*Source: Author's calculations.*

All sectoral indices are associated with an immediate positive reaction to a negative monetary policy shock (an increase of the reference interest rate), which may be opposite of what conventional economic theory predicts. It may imply that economic agents i.e., market participants interpret policy tightening as a positive signal in anticipation of reduction in inflationary pressures, hence improving their stock-market sentiment. This is particularly the case for the telecom and financial sectors, whereby the shock effect lives for about a year and is statistically significant. The delayed subsequent mean-reversion observed particularly in the telecom sector could be associated with its capital-intensive nature i.e., higher costs of financing which may impact profitability and therefore stock prices for which it may take time to fully materialize in such markets, provided the level of information-efficiency of the regional markets.

The response from the Finance sector is, however, logical from the perspective that a monetary policy shock is absorbed by this sector and transferred onto the final consumer resulting in interest margin expansion and improved profitability for banks, having in mind the bank-financed nature of the region. The mean reverting effect could be associated with changes in loan demand.

The sharper decline in the electricity sector, in contrast to the telecom sector, could be attributed to the demand. Initial perceptions of inelasticity may yield to behavioral adjustments in consumption. This suggests that market participants may subsequently modify their electricity needs, resulting in consumption reduction. Moreover, initiatives for green transition after the recent structural shocks (OECD, 2022) may imply that the electricity sector faces a need for more frequent technological upgrades resulting in high costs of financing. However, it is worth noting that the reaction of the electricity sector may not be statistically significant, which portrays the market reaction as semi-effective at best.

In a similar vein, the manufacturing sector has an initial positive response to a monetary policy shock, yet its mean reversion is much steeper and negative, stabilizing after approximately (6) months post shock. This raises the possibility of two potential factors i) higher borrowing costs coupled with ii) consumer spending constraints i.e., increase in interest expense and demand reduction take effect within month post policy shock. However, the response is not statistically significant.

Overall, the impulse response functions suggest that the regional stock markets react to monetary policy shocks only partially and to a limited magnitude. The response is positive only in the telecom and finance sectors, suggesting that the operability of the asset price channel of monetary policy when observed regionally is limitedly present likely due to the presence of multinational brands who serve integrative function of the local stock (and overall financial) markets. Responses' mean reversion in about a year could be due to i) the still semi-efficient nature of the regional markets i.e., slower assimilation of information by broader market participants, ii) limited liquidity of these markets, and iii) reliance on bank-based financing rather than equity financing.

### 4.3 Cointegration and long-run relationships

Table (3) presents the Kao panel test for co-integration of our included variables, considering one regional index at a time. All subtests are significant at 0.05 level. On that ground, we robustly reject the null hypothesis of no cointegration i.e., conclude that long-run co-movement exists among the variables. Subsequently, we proceed with a Panel VECM i.e., apply a PMG estimation method.

Table 3. Kao co-integration test.

| *Variable name* | *Variable abbreviation* | *p-values* |
|---|---|---|
| | *Ho: No co-integration among variables (p-values)* | |
| **Stock-market index of Telecom** | $indextelecom_{jt}$ | 0,0000 |
| **Stock-market index of Manufacturing** | $indexman_{jt}$ | 0,0000 |
| **Stock-market index of Electricity** | $indexelec_{jt}$ | 0,0000 |
| **Stock-market index of Finance and Insurance Activities** | $indexfin_{jt}$ | 0,0000 |
| *Source: Author's calculations.* | | |

**Table 4** presents the long-run coefficients obtained from the Pooled Mean Group (PMG) estimation. The results indicate statistically significant relationships between monetary-policy variables and sectoral stock indices, though the strength and sign of the effects vary across sectors. The error-correction terms are negative and statistically significant in all specifications, suggesting the presence of long-run cointegration and gradual convergence toward equilibrium following short-run deviations. The adjustment appears slow, with implied half-lives between 1.5 and 3 years, which is consistent with the limited liquidity and low turnover typical of small and segmented markets.

Table 4. Long-run relationships

| | **Stock-market indices per sector** | | | |
|---|---|---|---|---|
| **VARIABLES** | Manufacturing | Finance | Electricity | Telecom |
| | | | | |
| **Error correction term** | -0.0301** | -0.0486*** | -0.0549*** | -0.0383*** |
| | (0.014) | (0.014) | (0.002) | (0.014) |
| **Log of industrial production** | 0.643* | 0.568 | 0.628 | -0.386 |
| | (0.376) | (0.452) | (0.420) | (0.381) |
| **Log of CPI** | -0.773* | -0.15 | -0.682 | -0.631** |
| | (0.414) | (0.594) | (0.729) | (0.304) |
| **Changes in reserves** | -1.045*** | -1.278*** | -0.0717 | -0.0064 |
| | (0.158) | (0.133) | (0.177) | (0.154) |
| **Reference interest rate** | 3.887** | 9.725*** | 3.043 | 3.026*** |
| | (1.826) | (3.343) | (2.241) | (1.021) |

*Source: Author's calculations.*

*Note: *, ** and *** signify statistical significance of the coefficients at 10%, 5% and 1%, respectively. Standard errors are included in parentheses.*

Overall, our findings from the separation of the long- from the short-run dynamics suggest that the asset price channel plays some but not full role in the region. They corroborate the findings from the impulse responses based on a PVAR framework and suggest that the asset price channel of monetary policy is relevant at the regional level to the extent regional sub-markets are integrated, which is probably mostly spurred by the presence of multinational brands through their daughter-companies in the republics of former Yugoslavia. The opposite holds true as well: if sub-markets are manly isolated at the level of a country, for example because of being in dominant state ownership, like the electricity sector, or because their integration is not as advanced as in finance or telecom, like the manufacturing, then asset price channel may not be as effective. Despite, in the case of manufacturing, the result is mixed, which may also imply that further disaggregation of manufacturing may warrant investigation should richer datasets become available.

## 5. Conclusion

This study examines the operation of the asset price channel of monetary policy across six successor states of former Yugoslavia. The analysis constructs regional sectoral stock indices for manufacturing, finance, telecommunications, and electricity. The motivation stems from the idea that greater regional cooperation and exchange connectivity could improve market liquidity and enhance the transmission of monetary policy.

Empirical results from a Panel VAR and a Pooled Mean Group (PMG) estimation suggest a limited, though discernible, operation of the asset price channel in the region. Sectoral stock indices in finance and telecommunications respond to changes in monetary conditions, consistent with their stronger integration into cross-border financial networks and foreign ownership structures. In contrast, manufacturing and electricity sectors, characterized by state ownership and low market activity, show weaker responsiveness.

The negative but small error-correction terms imply that adjustments toward long-run equilibrium are slow, typically extending over several quarters. This pattern is consistent with structural features of regional markets, including shallow liquidity, narrow investor bases, and bank-dominated financial systems.

The specification accounts for both domestic and external monetary influences. The use of Euribor as a policy-rate proxy for euro-linked economies, together with central-bank foreign-exchange reserves as a measure of liquidity conditions, captures the dual nature of monetary-policy transmission in this hybrid institutional environment. Some countries, notably Slovenia and Bosnia and Herzegovina, are influenced by the Eurosystem's monetary stance, while non-Eurozone economies maintain national policy rates but remain exposed to external spillovers through trade, capital flows, and exchange-rate stabilization efforts.

A robustness check using the Mean Group (MG) estimator, which relaxes the assumption of long-run homogeneity, yields similar results for the finance sector but greater sensitivity for reserves and price variables. This supports the view that the asset price channel is partial, heterogeneous, and context-dependent rather than systematic across the region.

Taken together, these findings are consistent with the view that enhancing market liquidity through stronger regional cooperation including harmonized listing standards, improved information flows, and gradual institutional convergence may deepen financial intermediation and improve policy transmission over time. This interpretation is not advanced as a proven causal result, but as a plausible policy implication consistent with observed functional relationships and with current regulatory intentions toward regional market integration.

Future research could extend this analysis using a structural panel VAR framework to identify more precisely the channels through which monetary shocks propagate to sectoral asset prices. Such work would help verify whether the patterns observed here persist under stricter identification assumptions.

## 6. References

Abuka., C., Alinda., R., Jose-Lius., P. and Presbitero., A. (2015) Monetary Policy in a Developing Country: Loan Applications and Real Effects. IMF Working Paper

Ajaz., T., Nain, M. Z., Kamaiah, B., & Sharma, N. K. (2017) Stock prices, exchange rate and interest rate: evidence beyond symmetry. *Journal of Financial Economic Policy*, 9(1), pp. 2-19.

Altunbas, Y., Fazlov, O., and Molyneux, P. (2002) Evidence on the Bank Lending Channel in Europe. *Journal of Banking and Finance*, 26(11), pp. 2093-2110.

Andreasen., E., and Valenzuela, P. (2016) Financial openness, domestic financial development and credit raitings. *Finance Research Letters*, 16, pp. 11-18

Belmouss., F. El Ouazzani, Y. and Mafamane, D. (2023) Effectiveness of the asset price channel as a transmission mechanism for monetary policy in Morocco: Evidence from a VAR analysis, *International Journal of Accounting, Finance, Auditing, Management and Economics*, 4(3-2), pp. 716-741.

Berkman, N.G., 1978. On the significance of weekly changes in M1. *New England economic review*, *78*(1978), pp.5-22.

Bernanke., B, Gertler, M., and Gilchrist, S. (1996) The Financial Accelerator and the Flight to Quality. *The Review of Economics and Statistics*, 78(1), pp. 1–15.

Bernanke., B. and Blanchard, O. (2023) What Caused the U.S. Pandemic-Era Inflation? Hutchins Center Working Paper 86.

Bernanke., B. S. and Kuttner., K. N. (2005) What Explains the Stock Market's Reaction to Federal Reserve Policy? Mimeo, Board of Governors and Federal Reserve Bank of New York.

Bernanke., B. S., and Gertler, M. (1995) Inside the Black Box: The Credit Channel of Monetary Policy Transmission. *The Journal of Economic Perspectives*, 9(4), pp. 27–487.

Bernanke., Ben & Gertler, M. (1989) Agency Costs, Net Worth, and Business Fluctuations. *American Economic Review*, 79(1), pp. 14-31.

Bernanke., S. B. (2000) A Perspective on Inflation Targeting : Remarks before the Annual Washington Policy Conference, National Association of Business Economists, Washington, D.C.

Bondt., G.D. (1999) Banks and Monetary Transmission in Europe: Empirical Evidence. *BNL Quarterly Review*, 52(209), pp. 149-168.
Cassola., N. and Morana, C. (2004) Monetary Policy and the Stock Market in the Euro Area. *Journal of Policy Modeling*, 26, pp. 387-399.

Cecchetti., G. S. Genberg. H., Lipsky. J. and Wadhwani., S. (2000) Asset Prices and Central Bank Policy. The Geneva Report on the World Economy No. 2.

Chami., Cosimano F. T. and Fullenkamp, C. (1999) The Stock Market Channel of Monetary Policy. IMF Working Paper No. 99/22

Choi., S, Willems., T, and Yoo Y. S. (2022) Revisiting the Monetary Transmission Mechanism Through an Industry-Level Differential Approach. IMF Working paper 2022/017.

Cornell., B. (1983) The money supply announcements puzzle: review and interpretation. *American Economic Review*, 73, pp. 644-657.

Crowder., J. William (2006) The Interaction of Monetary Policy and Stock Returns. *The Journal of Financial Research,* 29(4), pp.523-535.

Dahl, J., Giudici, V., Senggupta, J., Kim, S., and Ervin Ng. (2019) Bracing for Consolidation: The Quest for Scale. *Asia-Pacific Banking Review* 2019, pp. 1-41.

Delgado., M and Gravelle. T. (2023) Central bank asset purchases in response to the Covid-19 crisis. BIS Papers, No. 68

Ehrmann., M, and Fratzscher, M. (2004) Taking Stock: Monetary Policy Transmission to Equity Markets. *Journal of Money, Credit and Banking*, 36(4), pp. 719-737.

Ehrmann., M., Gambacorta, L., Martinez Pagés, J., Sevestre, P., and Worms, A. (2001) Financial Systems and The Role of Banks in Monetary PolicyTransmission in the Euro Area. Working Paper Series, No 105, European Central Bank.

Ercegovac, R., Šestanović, T. and Pečarić, M., (2025) ECB quantitative tightening: Euribor-Overnight Index Swap spread and transmission mechanism efficiency. Bank i Kredyt, 56(2), pp.163-184.

Fama., E. F. (1970) Efficient capital markets: a review of theory and empirical work. *Journal of Finance*, 25, pp. 383-417.

Friedman., M. and Schwartz, A. (1963) Money and Business Cycles. *Review of Economics and Statistics*, 45, pp. 32-64.

Fuhrer., J.C. (1996) Monetary policy shifts and long-term interest rates. *Quarterly Journal of Economics*, 111(4), pp. 1183-1209

Gagliardone., L. and Gertler, M. (2023) Oil Prices, Monetary Policy and Inflation Surges. NBER Working Papers 31263.

Galí, Jordi, and Luca Gambetti. (2015) The Effects of Monetary Policy on Stock Market Bubbles: Some Evidence. *American Economic Journal: Macroeconomics*, 7(1), pp. 233-57

Gertler., S. and Gilchrist, S. (1994) Monetary Policy, Business Cycles, and the Behavior of Small Manufacturing Firms. *The Quarterly Journal of Economics*, 109(2), pp. 309 - 340

Grabowski, W., Stawasz-Grabowska, E. (2021) How have the European central bank's monetary policies been affecting financial markets in CEE-3 countries?. *Eurasian Econ Rev*, 11, pp.43-83

Guariglia., A., and Markose, S. (2000) Voluntary Contributions to Personal Pension Plans: Evidence from the British Household Panel Survey. *Fiscal Studies*, 21(4), pp. 469–488.

Hafer, R. W. (1986) The Response of Stock Prices to Changes in Weekly Money and the Discount Rate. Federal Reserv Bank of St. Louis Review, 68(3), pp. 5-14.

Hamburger., M. J. and Kochin, L. A. (1972) Money and stock prices: the channels of influence. *Journal of Finance*, 27, pp. 231-249.

Hardouvelis., A. G. (1987) Macroeconomic Information and Stock Prices. *Journal of Economics and Business*, 39, pp. 131-140.

Hsing, Y., & Hsieh, W-jen . (2014) Test of the Bank Lending Channel for a BRICS Country. *Asian Economic and Financial Review*, 4(8), pp. 1016–1023.
IMF (2023) Switzerland: ECB Monetary Policy Spillovers on Swiss Stock Market.

IMF (2021) Miles to go: The future of emerging markets – IMF F&D.

IMF (2023) Switzerland: ECB Monetary Policy Spillovers on Swiss Stock Market.

Jarocíński, M. (2010) Responses to Monetary Policy Shocks in The East and The West of Europe, A Comparison. ECB, Working Paper No. 970

Kashyap., A. and Stein, J. (2022) Monetary policy when the central bank shapes financial-market sentiment. NBER Working Paper Series.

Kashyap., K. A. and Stein., C. J. (1994) The impact of monetary policy on bank balance sheets. Carnegie-Rochester Conference Series on Public Policy, 42, pp. 151-195.

Keran., M. W. (1971) Expectations, money and the stock market. Federal Reserve Bank of St. Louis Review, pp. 16-31.

Kurniawan., A. and Dwi Astuti, R. (2023) "Transmission Mechanism of Monetary Policy Through Asset Price and Exchange Rate Channel in Indonesia", *JIET* (Jurnal Ilmu Ekonomi Terapan), 8(1), pp. 41–54.

Kuttner., K. N. (2001) Monetary policy surprises and interest rates: Evidence from the Fed funds futures market. *Journal of Monetary Economics*, 47(3), pp. 523-544.

Lastrapes, W. D. (2001) International Evidence on Equity Prices, Interest Rates, and Money. *Journal of International Money and Finance*, 17(3), pp.377-406.

Lee., C,. and Chou, I. (2018). *Finance Research Letters*, 25(C), pp.124-130.

Li., H. Ni, J. Xu. Y. and Zhan. M. (2021) Monetary policy and its transmission channels: Evidence from China. *Pacific-Basin Finance Journal*, 68(C).

Lynge., M. J. Jr. (1981) Money supply announcements and stock prices. The *Journal of Portfolio Management*, 8, pp. 40-43.

Matousek., R and Solomon, H. (2018) Bank lending channel and monetary policy in Nigeria. *Research in International Business and Finance*, 45, pp. 467-474.

Mehrotra., A. and Schanz, J. (2020) Financial Market development and monetary policy. BIS Papers, 113.

Mishkin., S. F. (1996) The Channels of Monetary Transmission: Lessons for Monetary Policy. NBER Working Papers 5464.

fOECD (2022) Latin America and the Caribbean: The green transition can be an economic and social game changer, says new report. OECD Library

Olivero., P. M., Yuan., L. and Jeon. N. B. (2011) Competition in banking and the lending channel: Evidence from bank-level data in Asia and Latin America. *Journal of Banking and Finance*, 35(3), pp. 560-571.

Patelis, A. D. (1997) Stock Return Predictability and The Role of Monetary Policy. *The Journal of Finance*, 52(5), pp. 1951–1972.

Pearce., K. D. and Roley., V. V. (1983) The Reaction of Stock Prices to Unanticipated Changes in Money; A Note. *The Journal of Finance*, 38(4), pp. 1323-1333.

Pearce. D and Roley. V. (1985) Stock Prices and Economic News. *The Journal of Business*, 58(1), pp.49-67

Pesando., J. E. (1974) The Supply of Money and Common Stock Prices: Further Observations on the Econometric Evidence. *The Journal of Finance*, 29(3), pp. 909–921.

Rigobon, R., and Sack, B. (2003) Measuring the Response of Monetary Policy to the Stock Market. *Quarterly Journal of Economics*, 118(2), pp. 639-669.

Rogalski., R. J. and Vinso, J. D. (1977) Stock Returns, Money Supply and the Direction of Causality. *Journal of Finance*, 32(4), pp. 1017-1030.

Roley, V. V. (1982) Weekly Money Supply Announcements and the Volatility of Short-term Interest Rates. *Economic Review*, Federal Reserve Bank of Kansas City, 67, pp. 3-15.

Roley., V. V. and Troll. R. (1984) The impact of discount rate changes on market interest rates. *Economic Review*, Federal Reserve Bank of Kansas City, 69, pp.27-39

Rozeff., S. M. (1974) Market efficiency and the lag effect of monetary policy. *Journal of Financial Economics*, 1, pp. 245-302.

Santomero, A. M. (1983) Controlling Monetary Aggregates: The Discount Window. *Journal of Finance*, 38(3), 827-843.

Sellin., P. (2001) Monetary Policy and the Stock Market: Theory and Empirical Evidence. *Journal of Economic Surveys*, 15(4), pp. 491-541.

Sims, C. A. (1980) Macroeconomics and Reality. *Econometrica*, 48(1), pp. 1–48.

Smirlock., M. and Yawitz, J. (1985) Asset Returns, Discount Rate Changes, and Market Efficiency. *The Journal of Finance*, 40(4), pp. 1141–1158.

Soedarmono., W. Gunadi., I., Pambudi., S. and Nurhayati., T. (2021) The Bank Lending Channel Revisited: Evidence From Indonesia. Bank of Indonesia, Working Paper 04/2021.

Sprinkel., B. W. (1964) Money and Stock Prices. Homewood, IL: Richard D. Irwin.

Stoica., O. Nucu, A. E., & Diaconasu, D.-E. (2014) Interest Rates and Stock Prices: Evidence from Central and Eastern European Markets. *Emerging Markets Finance & Trade*, 50, pp. 47–62.

Taylor., B. J. (1995) The Monetary Transmission Mechanism: An Empirical Framework. *Journal of Economic Perspectives*, 9, pp.11-26.

Tchereni, B.H.M., Makawa, A., & Banda, F. (2022) Effectiveness of the Asset Price Channel as a Monetary Policy Transmission Mechanism in Malawi: Evidence from Time Series Data. International *Journal of Economics and Financial Issues*, 12(5), pp. 160-168.

Thorbecke, W. (1997) On Stock Market Returns and Monetary Policy. *Journal of Finance*, 52(2), pp. 635-654.

Thorbecke, W. (2023) The Impact of Monetary Policy on the U.S. Stock Market since the Pandemic. Discussion papers 23054, Research Institute of Economy, Trade and Industry (RIETI).

Tobin, James (1969). "A General Equilibrium Approach to Monetary Theory." *Journal of Money, Credit and Banking* , 1, pp. 15-29.

Tongurai, J., and Vithessonthi, C. (2023) Financial openness and financial market development. *Journal of Multinational Financial Management*, 67(C).

Urich, T. and Wachtel, P. (1981) Market Response to the Weekly Money Supply Announcements in the 1970s. *Journal of Finance, American Finance Association*, 36(5), pp. 1063-1072.

Waud., R. N. (1970) Public interpretation of federal reserve discount rate changes: evidence on the `announcement effect'. *Econometrica*, 38, pp. 231-250.

Woolrdige, P. (2020) Implications of financial market development for financial stability in emerging market economies. BIS Paper.

Naoyuki., Y. Farhad. H.T. Hassanzadeh. F. Prasetyo. A. and Danu. A. (2014) Response of Stock Markets to Monetary Policy: An Asian Stock Market Perspective. ADBI Working Paper No. 497

Zhan, S., Tang, Y., Li, S., Yao, Y., & Zhan, M. (2021). How does the money market development impact the bank lending channel of emerging Countries? A case from China. *The North American Journal of Economics and Finance*, 57(C)

## Appendix 1

### Sectoral classification according to NAC.REV2

Table A1

| | Sector: Electricity, gas, steam and air conditioning supply |
|---|---|
| 1 | Elektroprivreda Crne Gore AD |
| 2 | Elektroprivreda BiH d.d. Sarajevo |
| 3 | Elektroprivreda HZHB Mostar d.d. |
| 4 | Crnogorski Elektroprenosni Sistem AD |
| 5 | Fintel Energija AD |

Table A2

| | Sector: Financial and insurance activities |
|---|---|
| 1 | Stopanska Banka AD |
| 2 | TTK Banka AD |
| 3 | UNI Banka AD |
| 4 | NLB Banka AD |
| 5 | Komercijalna Banka AD |
| 6 | Hipotekarna Banka d.d. |
| 7 | NLB Slovenia d.d. |
| 8 | Asa Banka d.d. Sarajevo |
| 9 | Hrvatska Postenska Banka d.d. |
| 10 | Dunav Osiguranje AD |
| 11 | Triglav Osiguranje d.d |
| 12 | SAVA Group d.d. |

Table A3

| | Sector: Information and Telecommunications |
|---|---|
| 1 | JP Hrvatske Telekomunikacije d.d. Mostar |
| 2 | Makedonski Telekom AD |
| 3 | Telekom Slovenia d.d |
| 4 | BH Telecom Sarajevo d.d |
| 5 | Crnogorski Telekom d.d. |
| 6 | Hrvatski Telekom d.d. |

Table A4

| | **Sector: Manufacturing** |
|---|---|
| 1 | Podravka d.d. |
| 2 | Atlantic d.d. |
| 3 | Plantaze Podgorica AD |
| 4 | Impol Seval d.d. |
| 5 | Badeco Adria d.d. |
| 6 | Messer Tehnogas d.d. |
| 7 | Cinkarna Celje d.d. |
| 8 | Jedinstvo Sevojna d.d. |
| 9 | KRKA d.d. |
| 10 | Bosnija Ljek d.d. |
| 11 | Alkaloid AD |
| 12 | Rade Koncar d.d. |
| 13 | Energoinvest d.d. |

**Appendix 2**

To assess the robustness of the baseline PMG results, the long-run relationships were re-estimated using the Mean Group (MG) estimator, which relaxes the homogeneity restriction across countries.

The MG estimates (Table A2.1) confirm a positive and significant association between real activity and the finance-sector index, while the coefficients on prices, foreign-exchange reserves, and the reference interest rate exhibit variation in sign and significance across sectors.

The finance sector remains the most responsive to monetary-policy conditions, but the but the **loss of significance** in the reserves coefficient relative to the PMG model highlights the role of cross-country heterogeneity in external-balance dynamics.

Given the short time span of the panel and the institutional diversity of the sample (euroized vs. non-euroized regimes), these differences are expected. Broadly, the MG results support the qualitative findings of the PMG model while indicating that some long-run elasticities are not invariant to estimator choice and should therefore be interpreted with caution.

**Table A2.1 Mean Group (MG) Long-Run Coefficients by Sector**

| | Stock-market indices per sector | | | |
|---|---|---|---|---|
| **VARIABLES** | **Manufacturing** | **Finance** | **Electricity** | **Telecom** |
| **Log of industrial production** | 0.031 | 0.498* | -0.050 | -0.224 |
| | (0.212) | (0.247) | (0.052) | (0.146) |
| **Log of CPI** | -1.476 | -1.145 | -3.042 | -1.862* |
| | (1.366) | (0.988) | -1.685 | (0.798) |
| **Changes in reserves** | -0.724*** | 1.052 | 0.467 | -0.097 |
| | (0.163) | (0.543) | (0.441) | (0.259) |
| **Reference Interest rate** | 0.035*** | -0.023 | -0.008 | 0.009 |
| | (0.003) | (0.062) | (0.022) | (0.019) |

*Source: Author's calculations.*

*Note: *, ** and *** signify statistical significance of the coefficients at 10%, 5% and 1%, respectively. Standard errors are included in parentheses.*